\newif\ifanon
\newif\ifsubmission
\submissiontrue

\ifsubmission
  \documentclass[runningheads]{llncs}
  \providecommand{\keywords}[1]{\par\addvspace\baselineskip
  \noindent{\bf Keywords:}\enspace\ignorespaces#1}%
  \date{}
  \anontrue
  \author{Hamza Jeljeli}
  \institute{CARAMEL project-team, LORIA, INRIA / CNRS / Universit\'e de Lorraine,\\
           Campus Scientifique, BP 239,
           54506 Vand\oe{}uvre-l\`es-Nancy Cedex, France\\
           \url{Hamza.Jeljeli@loria.fr}
  }
  \makeatletter
  \renewcommand\subsubsection{\@startsection{subsubsection}{3}{\z@}%
    {-18\p@ \@plus -4\p@ \@minus -4\p@}%
    {-0.5em \@plus -0.22em \@minus -0.1em}%
    {\normalfont\normalsize\bfseries\boldmath}}
  \makeatother
\else
  \documentclass[runningheads]{llncs}
  \anonfalse
\fi

\usepackage[utf8]{inputenc}
\usepackage[english]{babel}
\usepackage[T1]{fontenc}
\usepackage{amsmath,amssymb}
\usepackage{comment}
\usepackage{hyperref}
\usepackage{multirow}
\usepackage{url}
\usepackage{pgf}
\usepackage{subfig}

\usepackage{tikz} 
\usetikzlibrary{arrows}
\usetikzlibrary{snakes}
\usepackage{pgffor,xifthen}
\usepackage{pgfplots}
\usetikzlibrary{patterns}
\usepackage{siunitx}
\usepackage{numprint}
\usepackage{bm}

\definecolor{violet}{rgb}{0.5,0,0.5}

\title{Resolution of Linear Algebra for the Discrete Logarithm
Problem Using GPU and Multi-core Architectures}
\titlerunning{Resolution of Linear Algebra for DLP Using Parallel Architectures}

\ifanon\else
\author{Hamza Jeljeli}
\authorrunning{H. Jeljeli}
\institute{CARAMEL project-team, LORIA, INRIA / CNRS / Université de Lorraine,\\
           Campus Scientifique, BP 239,
           54506 Vandœuvre-lès-Nancy Cedex, France\\
           \url{Hamza.Jeljeli@loria.fr}}
\fi

\begin{document}

\maketitle

\begin{abstract}
    In cryptanalysis, solving the discrete logarithm problem (DLP) is key to assessing the security of many public-key cryptosystems. The index-calculus methods, that attack the DLP in multiplicative subgroups of finite fields, require solving large sparse systems of linear equations modulo large primes. This article deals with how we can run this computation on GPU- and multi-core-based clusters, featuring InfiniBand networking. More specifically, we present the sparse linear algebra algorithms that are proposed in the literature, in particular the block Wiedemann algorithm. We discuss the parallelization of the central matrix--vector product operation from both algorithmic and practical points of view, and illustrate how our approach has contributed to the recent record-sized DLP computation in GF($2^{809}$).
    
\end{abstract}



\section{Introduction}
\vspace*{-0.25cm}
The security of several public-key cryptosystems and protocols relies on the hardness of the computation of the discrete logarithm problem (DLP) in a given cyclic group~\cite{ODLY84}. To name but a few, we can mention the Diffie–Hellman key exchange protocol \cite{DH76}, the ElGamal encryption system \cite{ElGamal85} or the pairing-based cryptography \cite{Pairing}.

In this context, a family of algorithms, known as \textit{index-calculus} methods, is used to attack the DLP on finite fields. The majority of these algorithms propose to solve it in time sub-exponential in the size of the finite field. While a stream of recent algorithmic improvements for fields of small characteristic, including a quasi-polynomial algorithm~\cite{QPA}, have produced several record-sized computations~\cite{GRAN14}, the sub-exponential methods appear to be most competitive for fields of prime extension degree, at least so far.  

Index calculus algorithms require solving large sparse systems of linear equations over finite fields. It is important to mention that,
 most considerations and methods in the case of numerical computations do not apply here. Several papers have focused on efficient implementations of sparse linear algebra over finite fields. For instance, Schmidt et al.~\cite{SHMI11} treated linear algebra over GF(2) for integer factorization; Boyer et al.~\cite{BOYE10} worked on the case of small finite rings and fields.

\vspace*{-0.5cm}
   
\subsubsection{Problem Statement.}
Let GF($q$) be the field in which the DLP is to be solved. The linear algebra is performed modulo a large prime $\ell$ that divides $q-1$. We consider $\ell$ between 160 and 650 bits, along with an $N$-by-$N$ sparse matrix $A$ defined over $\mathbb{Z}/\ell\mathbb{Z}$. The size $N$ ranges from hundreds of thousands to millions. Each row of $A$ contains $O\left((\log N)^2\right)$ non-zero coefficients. The very first columns of $A$ are relatively dense, then the column density decreases gradually. The row density does not change significantly (cf. Figure~\ref{fig:balancing}). The so-called linear algebra step in the DLP computation consists in finding a non-trivial vector $w \in (\mathbb{Z}/\ell\mathbb{Z})^N$ such that $A w =0$.  

We assume that we have access to one or several high-performance computing clusters, containing multi-core CPUs and/or GPUs, interconnected by fast communication links (typically InfiniBand). We want to optimize the use of these resources in order to solve the linear algebra problem efficiently. In particular, we aim to minimize the overall wall-clock time for solving the problem. First, at an algorithmic level, we study how these heavy computations can be distributed into smaller parallel subtasks. Then, we focus on more practical concerns, for instance the communication within these different subtasks.     

\vspace*{-0.5cm}

\subsubsection{Organization.}
This article is organized as follows: Section~\ref{sec:wied} gives an overview of the relevant algorithms for sparse linear algebra, while we discuss the parallelization of the matrix--vector product operation and focus on the communication concerns in Section~\ref{sec:mat-vec}. Finally, Section~\ref{sec:examples of computation} details how our implementation has been used in concrete DLP computations with different hardware setups.  
 
\vspace*{-0.25cm}
 
\section{Algorithms for Sparse Linear Algebra}
\vspace*{-0.25cm}

To solve systems of linear equations, two families of algorithms are available: \textit{direct methods}, such as Gaussian elimination or LU/QR decompositions, and \textit{iterative methods}, such as the conjugate gradient method and, in the context of linear algebra over finite fields, the Lanczos~\cite{LANC52} and Wiedemann~\cite{WIED86} algorithms.

The first set of algorithms requires $O\left(N^\omega\right)$ field operations, where $\omega$ is, for implementation concerns, 2.81 at best using the Strassen algorithm for matrix multiplication~\cite{STRA69}. However, these methods tend to densify the matrix, which quickly raises storage issues. The second set of algorithms does not modify the matrix and requires $O\left(N\right)$ sparse-matrix--vector products (SpMVs). As long as an SpMV can be performed faster than $O\left(N^{\omega - 1}\right)$ field  operations, the iterative methods are asymptotically faster. This condition is reasonable, since the complexity of an SpMV is $O\left(N\gamma\right)$, where $\gamma$ is the average number of non-zero coefficients per row. From both storage and complexity points of view, the iterative methods appear to be more suited to sparse linear algebra.

Still, in our case, despite the fact that the matrix is extremely sparse, the cost of an iterative solver remains high because the matrix is very large. The exact nature of the computation calls for no less than  $N$ iterations, or a number proportional to $N$ depending on some fine points. The approach following is applied to tackle that problem. First, a structured Gaussian elimination (SGE) is run as a preprocessing step so as to reduce the size of the matrix~\cite{POME92}; then an iterative solver is used. Although the Gaussian elimination increases the average row weight, it nevertheless allows us to decrease the cost of the iterative solver and to reduce the amount of required memory, which is a major implementation concern as will be seen in the following. It is important that we stop the SGE when the projected cost of the iterative solver starts to increase again or when memory requirements are small enough so as to fit on the hardware at hand~\cite{Bouvier12}.
 
The Lanczos and Wiedemann algorithms are the most commonly used iterative algorithms in the context of finite fields linear algebra. The Lanczos algorithm is known to have a better complexity than the Wiedemann algorithm. However, the block extension of Wiedemann algorithm (\textit{a.k.a} block Wiedemann) offers the opportunity to split the computation into several independent subtasks, which is an important practical advantage~\cite{KLEI10,KAZU07}.

The Wiedemann algorithm and block Wiedemann algorithms return both a vector $w$ of the kernel of $A$. This vector is non-trivial with high probability. In practice, a single  run of the solver is sufficient to find an appropriate solution. 

\label{sec:wied}

\vspace*{-0.25cm}

\subsubsection{Wiedemann algorithm}

The starting point of the Wiedemann algorithm is to choose two random vectors $x,y \in (\mathbb{Z}/\ell\mathbb{Z})^{N}$. The algorithm is organized in three steps~\cite{WIED86}, for which we use monikers borrowed from the CADO-NFS software implementation~\cite{cadonfs}.
\vspace*{-0.15cm}
\begin{itemize}
\item The first step computes the first $2N$ terms of the linearly recurrent sequence $\left (a_i\right )_{i \in \mathbb{N}} \in (\mathbb{Z}/\ell\mathbb{Z})^{\mathbb{N}}$, where $a_i = {}^txA^iy$. This step is usually called \textit{Krylov} .
\item Then, thanks to the Berlekamp--Massey algorithm, we compute the minimal polynomial of the sequence, which is the polynomial $F(X)=\sum_{i=0}^{d}f_iX^i$ of lowest degree $d$ such that $\sum_{i=0}^{d}f_ia_{k+i}=0$ for all $k\geq0$. The degree $d$ is close to $N$. We commonly call this step \textit{Lingen} . 
\item The last step, called \textit{Mksol} , finally computes $w=F(A)y$.
\end{itemize}
\vspace*{-0.15cm}

The Wiedemann algorithm requires $3N$ SpMVs for the \textit{Krylov} and \textit{Mksol} steps and $O\left( N \log N\right)$ field operations for the \textit{Lingen} step. 

\vspace*{-0.25cm}

\subsubsection{Block Wiedemann algorithm}
Wiedemann algorithm is fully sequential. In~\cite{KALT95,COPP94}, Coppersmith et al. presented a block variant that provides parallelism. The block Wiedemann algorithm replaces the vector $y \in (\mathbb{Z}/\ell\mathbb{Z})^{N}$ by a block of $n$ vectors $y^{(0)}, \dots, y^{(n-1)}$, each in $(\mathbb{Z}/\ell\mathbb{Z})^{N}$, and similarly uses a block of $m$ vectors for $x$. The sequence of scalars $a_i$ is thus replaced by a sequence of $m$-by-$n$ matrices. There is a complete freedom in the choice of the blocking parameters $(m,n)$. For the efficiency of the \textit{Lingen} run, $m$ is chosen to be equal to $2n$~\cite{cadonfs}. 
\vspace*{-0.2cm}

\begin{itemize}
\item The \textit{Krylov} step now computes the first $\left \lceil \frac{N}{n} \right 
\rceil + \left \lceil \frac{N}{m} \right \rceil$ terms of the 
sequence $\left (a_i \right) _{i \in \mathbb{N}}$. Notice that the $j$-th column of the $m$-by-$n$ matrix $^txA^iy$ depends only on the $j$-th column of the block vector $y$. Thus, the computation of $\left (^txA^iy\right )_{i \in \mathbb{N}}$ can be distributed into $n$ 
parallel tasks, each computing \linebreak $\left (^txA^iy^{(j)}\right )_{i \in \mathbb{N}}$. These tasks need no synchronization nor communication, except at the end when all their results are combined.

\item The \textit{Lingen} step seeks a linear generator for the previous sequence. The complexity of this step becomes $O\left(n^{\omega-1} N \log N\right)$ with $m=2n=o\left(\log N\right)$~\cite{THOME02}. The output of \textit{Lingen} is composed of $n$ generators $F^{(0)}, \dots, F^{(n-1)}$, each of them a polynomial over $\mathbb{Z}/\ell\mathbb{Z}$ of degree less than $\left \lceil \frac{N}{n} \right \rceil$.

\item The \textit{Mksol} step computes the following element of the null-space of $A$: $w= \sum_{j=1}^{n} F^{(j)}(A)y^{(j)}$. Similarly to the \textit{Krylov} phase, the computation can be distributed into $n$ independent computations.
\end{itemize}
\vspace*{-0.2cm}

In the rest of the paper, we focus on the \textit{Krylov} and \textit{Mksol} steps, as they dominate the overall cost and can benefit from parallel hardware. For the \textit{Lingen}  computation, we use the CADO-NFS software~\cite{cadonfs}.


\vspace*{-0.25cm}

\section{The Matrix--Vector Product}
\label{sec:mat-vec}
\vspace*{-0.25cm}

The \textit{Lingen} step complexity depends roughly quadratically on the blocking parameter $n$. Therefore, we can not increase too much the blocking parameters $(n,m)$. We observe also that the block Wiedemann algorithm does not distribute the matrix--vector product, so it does not reduce the amount of required memory per node. 
Thus, the parallelism provided by the block Wiedemann algorithm is soon limited. 
We need to explore how to carry out a \textit{Krylov}/\textit{Mksol} task on more than one computation node. Typically, this is related to performing each matrix--vector product in parallel on many computation nodes. In this section, we study how to accelerate this major operation on parallel hardware.


We assume that we have a set of identical \textit{computing nodes} organized according to a 2D rectangular grid and interconnected by a network. Each node is identified by its coordinates $(i,j)$ in the grid. At this level, we ignore the nature of the nodes. The nodes could be cores within a machine, independent machines or GPUs. The matrix $A$ is split into square parts of equal size, such that each node $(i,j)$ gets the part $A_{i,j}$.

\vspace*{-0.25cm}

\subsection{Communication/Computation scheme}
\vspace*{-0.25cm}

An SpMV iteration takes an input vector $u$ and computes $v=Au$. At the beginning of an iteration, a node $(i,j)$ holds the sub-matrix $A_{ij}$ and the $j$-th fragment $u_j$ of the input vector $u$. The nodes collaborate together to compute the output vector, which will be the input vector to the next iteration. To be able to run the next iteration, the node $A_{ij}$ only needs to know the $j$-th fragment $v_j$ of the output vector $v$. More specifically, the parallel SpMV product is performed as follows.
\vspace*{-0.25cm}

\begin{enumerate}
\item Each node $(i,j)$ computes the partial SpMV $A_{ij}u_j$.
\item Each diagonal node $(i,i)$ collects and sums the partial results from the nodes of the row~$i$. The sum corresponds to the $i$-th fragment of $v$.
\item Each diagonal node $(i,i)$ broadcasts its fragment $v_j$ to the nodes of the column~$i$.
\end{enumerate}
\vspace*{-0.25cm}

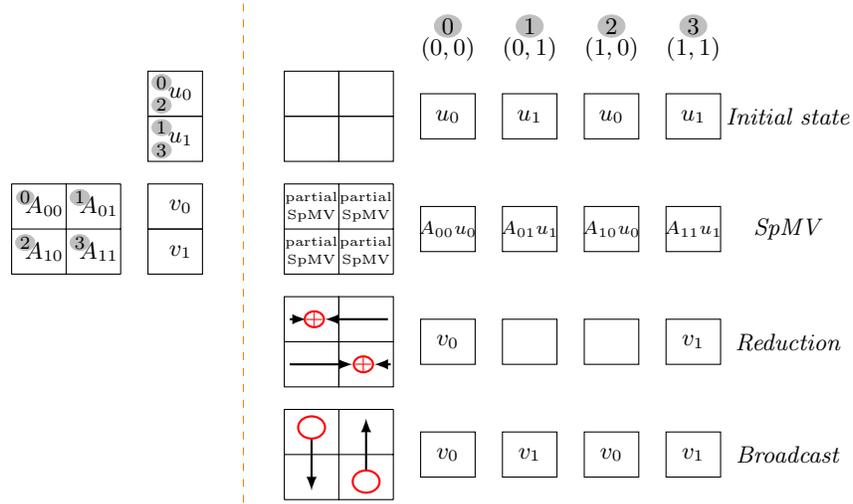
\begin{figure}[]
\centering
\begin{tikzpicture}[yscale=0.6, xscale=0.725]
\draw (0,0) grid (2,2);
\draw (0.6,1.5) node {$A_{00}$};
\draw[draw=gray!50, fill=gray!50] (0.25,1.7) circle (4.5pt);
\draw[black] (0.25,1.7) node {\scriptsize $0$};
\draw (1.6,1.5) node {$A_{01}$};
\draw[draw=gray!50, fill=gray!50] (1.25,1.7) circle (5pt);
\draw[black] (1.25,1.7) node {\scriptsize $1$};
\draw (0.6,0.5) node {$A_{10}$};
\draw[draw=gray!50, fill=gray!50] (0.25,0.7) circle (5pt);
\draw[black] (0.25,0.7) node {\scriptsize $2$};
\draw (1.6,0.5) node {$A_{11}$};
\draw[draw=gray!50, fill=gray!50] (1.25,0.7) circle (5pt);
\draw[black] (1.25,0.7) node {\scriptsize $3$};
\draw (2.5,2.5) rectangle (3.5,3.5);
\draw (2.5,3.5) rectangle (3.5,4.5);
\draw (3.1,4) node {$u_{0}$};
\draw[draw=gray!50, fill=gray!50] (2.75,4.25) circle (5pt);
\draw[black] (2.75,4.25) node {\scriptsize $0$};
\draw[draw=gray!50, fill=gray!50] (2.75,3.75) circle (5pt);
\draw[black] (2.75,3.75) node {\scriptsize $2$};
\draw (3.1,3) node {$u_{1}$};
\draw[draw=gray!50, fill=gray!50] (2.75,3.25) circle (5pt);
\draw[black] (2.75,3.25) node {\scriptsize $1$};
\draw[draw=gray!50, fill=gray!50] (2.75,2.75) circle (5pt);
\draw[black] (2.75,2.75) node {\scriptsize $3$};
\draw (2.5,0) rectangle (3.5,1);
\draw (2.5,1) rectangle (3.5,2);
\draw (3.1,1.5) node {$v_{0}$};
\draw (3.1,0.5) node {$v_{1}$};

\draw (5,4.5) rectangle (6,3.5);
\draw (6,4.5) rectangle (7,3.5);
\draw (5,3.5) rectangle (6,2.5);
\draw (6,3.5) rectangle (7,2.5);

\draw(7.5,3) rectangle (8.5,4);
\draw (9,3) rectangle (10,4);
\draw(10.5,3) rectangle (11.5,4);
\draw (12,3) rectangle (13,4);
\draw (8,3.5) node {$u_{0}$};
\draw (9.5,3.5) node {$u_{1}$};
\draw (11,3.5) node {$u_{0}$};
\draw (12.5,3.5) node {$u_{1}$};

\draw(5,0) grid (7,2);
\draw (5.5,1.3) node {\tiny SpMV};
\draw (5.5,1.7) node {\tiny partial};
\draw (6.5,1.3) node {\tiny SpMV};
\draw (6.5,1.7) node {\tiny partial};
\draw (5.5,0.7) node {\tiny partial};
\draw (5.5,0.3) node {\tiny SpMV};
\draw (6.5,0.7) node {\tiny partial};
\draw (6.5,0.3) node {\tiny SpMV};

\draw[draw=gray!50, fill=gray!50] (8,5.5) circle (7pt);
\draw[black] (8,5.5) node {$0$};
\draw[black] (8,5) node {$(0,0)$};
\draw[draw=gray!50, fill=gray!50] (9.5,5.5) circle (7pt);
\draw[black] (9.5,5.5) node {$1$};
\draw[black] (9.5,5) node {$(0,1)$};
\draw[draw=gray!50, fill=gray!50] (11,5.5) circle (7pt);
\draw[black] (11,5.5) node {$2$};
\draw[black] (11,5) node {$(1,0)$};
\draw[draw=gray!50, fill=gray!50] (12.5,5.5) circle (7pt);
\draw[black] (12.5,5.5) node {$3$};
\draw[black] (12.5,5) node {$(1,1)$};

\draw(7.5,0.5) rectangle (8.5,1.5);
\draw (9,0.5) rectangle (10,1.5);
\draw(10.5,0.5) rectangle (11.5,1.5);
\draw (12,0.5) rectangle (13,1.5);
\draw (8,1) node {\scriptsize $A_{00}u_{0}$};
\draw (9.5,1) node {\scriptsize $A_{01}u_{1}$};
\draw (11,1) node {\scriptsize $A_{10}u_{0}$};
\draw (12.5,1) node {\scriptsize $A_{11}u_{1}$};


\draw (5,-0.5) rectangle (6,-1.5);
\draw (6,-0.5) rectangle (7,-1.5);
\draw (5,-1.5) rectangle (6,-2.5);
\draw (6,-1.5) rectangle (7,-2.5);
\path[draw,->,>=latex,thick] (6.9,-1) -- (5.75,-1);
\path[draw,->,>=latex,thick] (5.1,-1) -- (5.4,-1);
\draw[draw=red, fill=white,thick] (5.55,-1) circle (5pt);
\draw[red] (5.55,-1) node {$+$};
\path[draw,<-,>=latex,thick] (6.25,-2) -- (5.1,-2);
\path[draw,<-,>=latex,thick] (6.65,-2) -- (6.95,-2);
\draw[draw=red, fill=white,thick] (6.45,-2) circle (5pt);
\draw[red] (6.45,-2) node {$+$};

\draw (7.5,-2) rectangle (8.5,-1);
\draw (9,-2) rectangle (10,-1);
\draw (10.5,-2) rectangle (11.5,-1);
\draw (12,-2) rectangle (13,-1);
\draw (8,-1.5) node {$v_{0}$};
\draw (12.5,-1.5) node {$v_{1}$};


\draw(5,-5) grid (7,-3);
\path[draw=black,->,>=latex, thick] (5.5,-3.3) -- (5.5,-4.8);
\draw[draw=red, fill=white,thick] (5.5,-3.4) circle (7pt);
\path[draw=black,<-,>=latex, thick] (6.5,-3.2) -- (6.5,-4.7);
\draw[draw=red, fill=white, thick] (6.5,-4.6) circle (7pt);

\draw(7.5,-4.5) rectangle (8.5,-3.5);
\draw (9,-4.5) rectangle (10,-3.5);
\draw(10.5,-4.5) rectangle (11.5,-3.5);
\draw (12,-4.5) rectangle (13,-3.5);
\draw (8,-4) node {$v_{0}$};
\draw (9.5,-4) node {$v_{1}$};
\draw (11,-4) node {$v_{0}$};
\draw (12.5,-4) node {$v_{1}$};


\draw [orange, dashed] (4.25,6) -- (4.25,-5.25);

\draw (14.25,3.5) node {\textit{Initial state}};
\draw (14.25,1) node {\textit{SpMV}};
\draw (14.25,-1.5) node {\textit{Reduction}};
\draw (14.25,-4) node {\textit{Broadcast}};

\end{tikzpicture}
\caption{Computation/Communication scheme for a $2 \times 2$ split of $A$}
\label{fig:2x2 split of A}
\vspace*{-0.75cm}

\end{figure}

In Figure~\ref{fig:2x2 split of A}, we give an example of a run for 4 parallel nodes with a $2 \times 2$ split of the matrix. In this figure, the 4 nodes are, represented in gray, numbered from 0 to 3. On the left-hand side, we indicate how the matrix $A$ and the input vector $u$ are distributed among the nodes. We detail on the right-hand side the intermediate data present on each node after each step.


The communication scheme suffers from the fact that only one node per row collects the partial products. A parallelization of the Reduction/Broadcast operations is possible, typically using the ReduceScatter/AllGather operations. This should yield to a significant speedup of the communication delay. However, the output of the iteration will be permuted, \textit{i.e.,} the fragments of $v$ will not be distributed as were those of $u$ in the beginning of the iteration. In summary, it remains an improvement that can be explored. 

\vspace*{-0.25cm}

\subsection{Balancing the workload}
\vspace*{-0.25cm}

The particular distribution of the non-zero coefficients is such that the nodes will get unbalanced workloads, and the nodes working on the denser parts will take more time than those working on the sparser ones. For the particular kind of input, this unbalance problem can fortunately be solved efficiently. To fix this problem, we apply permutations of the rows and columns, so that the distribution of non-zero coefficients for each sub-matrix is close to that of the matrix $A$, as shown in Figure~\ref{fig:balancing}. One possibility to obtain this permutation is to sort the columns by their weight and distribute them evenly among the nodes, then proceed likewise with the rows. This is made possible by the fact that the standard deviation of the row weight is much smaller than that of the column weight.


\begin{figure}[]
\centering

\begin{tikzpicture}[scale=0.7]

\begin{pgfscope}

\color{black}
\pgfputat{\pgfxy(-0.5,0.3)}{\pgfbox[left, center]{\textbf{Initial}}}
\pgfputat{\pgfxy(5.2,0.3)}{\pgfbox[left, center]{\textbf{Balanced}}}

\pgfdeclarehorizontalshading{myshadingD}
{5.5pt}{rgb(0pt)=(0,0,0); rgb(5pt)=(0.7,0.7,0.7); rgb(100pt)=(1,1,1)}
\pgftext[at=\pgfpoint{0cm}{0cm}] {\pgfuseshading{myshadingD}}

\pgfdeclarehorizontalshading{myshadingD}
{5.5pt}{rgb(0pt)=(0,0,0); rgb(6pt)=(0.7,0.7,0.7 ); rgb(100pt)=(1,1,1)}
\pgftext[at=\pgfpoint{0cm}{-5pt}] {\pgfuseshading{myshadingD}}

\pgfdeclarehorizontalshading{myshadingD}
{5.5pt}{rgb(0pt)=(0,0,0); rgb(7pt)=(0.7,0.7,0.7 ); rgb(100pt)=(1,1,1)}
\pgftext[at=\pgfpoint{0cm}{-10pt}] {\pgfuseshading{myshadingD}}

\pgfdeclarehorizontalshading{myshadingD}
{5.5pt}{rgb(0pt)=(0,0,0); rgb(9pt)=(0.7,0.7,0.7 ); rgb(100pt)=(1,1,1)}
\pgftext[at=\pgfpoint{0cm}{-15pt}] {\pgfuseshading{myshadingD}}

\pgfdeclarehorizontalshading{myshadingD}
{5.5pt}{rgb(0pt)=(0,0,0); rgb(7.5pt)=(0.7,0.7,0.7 ); rgb(100pt)=(1,1,1)}
\pgftext[at=\pgfpoint{0cm}{-20pt}] {\pgfuseshading{myshadingD}}

\pgfdeclarehorizontalshading{myshadingD}
{5.5pt}{rgb(0pt)=(0,0,0); rgb(6pt)=(0.7,0.7,0.7 ); rgb(100pt)=(1,1,1)}
\pgftext[at=\pgfpoint{0cm}{-25pt}] {\pgfuseshading{myshadingD}}

\pgfdeclarehorizontalshading{myshadingD}
{5.5pt}{rgb(0pt)=(0,0,0); rgb(7.5pt)=(0.7,0.7,0.7 ); rgb(100pt)=(1,1,1)}
\pgftext[at=\pgfpoint{0cm}{-30pt}] {\pgfuseshading{myshadingD}}

\pgfdeclarehorizontalshading{myshadingD}
{5.5pt}{rgb(0pt)=(0,0,0); rgb(8pt)=(0.7,0.7,0.7 ); rgb(100pt)=(1,1,1)}
\pgftext[at=\pgfpoint{0cm}{-35pt}] {\pgfuseshading{myshadingD}}

\pgfdeclarehorizontalshading{myshadingD}
{5.5pt}{rgb(0pt)=(0,0,0); rgb(10pt)=(0.7,0.7,0.7 ); rgb(100pt)=(1,1,1)}
\pgftext[at=\pgfpoint{0cm}{-40pt}] {\pgfuseshading{myshadingD}}

\pgfdeclarehorizontalshading{myshadingD}
{5.5pt}{rgb(0pt)=(0,0,0); rgb(11pt)=(0.7,0.7,0.7 ); rgb(100pt)=(1,1,1)}
\pgftext[at=\pgfpoint{0cm}{-45pt}] {\pgfuseshading{myshadingD}}

\pgfdeclarehorizontalshading{myshadingD}
{5.5pt}{rgb(0pt)=(0,0,0); rgb(13pt)=(0.7,0.7,0.7 ); rgb(100pt)=(1,1,1)}
\pgftext[at=\pgfpoint{0cm}{-50pt}] {\pgfuseshading{myshadingD}}

\pgfdeclarehorizontalshading{myshadingD}
{5.5pt}{rgb(0pt)=(0,0,0); rgb(14pt)=(0.7,0.7,0.7 ); rgb(100pt)=(1,1,1)}
\pgftext[at=\pgfpoint{0cm}{-55pt}] {\pgfuseshading{myshadingD}}

\pgfdeclarehorizontalshading{myshadingD}
{5.5pt}{rgb(0pt)=(0,0,0); rgb(15pt)=(0.7,0.7,0.7 ); rgb(100pt)=(1,1,1)}
\pgftext[at=\pgfpoint{0cm}{-60pt}] {\pgfuseshading{myshadingD}}

\pgfdeclarehorizontalshading{myshadingD}
{5.5pt}{rgb(0pt)=(0,0,0); rgb(13pt)=(0.7,0.7,0.7 ); rgb(100pt)=(1,1,1)}
\pgftext[at=\pgfpoint{0cm}{-65pt}] {\pgfuseshading{myshadingD}}

\pgfdeclarehorizontalshading{myshadingD}
{5.5pt}{rgb(0pt)=(0,0,0); rgb(14pt)=(0.7,0.7,0.7 ); rgb(100pt)=(1,1,1)}
\pgftext[at=\pgfpoint{0cm}{-70pt}] {\pgfuseshading{myshadingD}}

\pgfdeclarehorizontalshading{myshadingD}
{5.5pt}{rgb(0pt)=(0,0,0); rgb(15.5pt)=(0.7,0.7,0.7 ); rgb(100pt)=(1,1,1)}
\pgftext[at=\pgfpoint{0cm}{-75pt}] {\pgfuseshading{myshadingD}}

\pgfdeclarehorizontalshading{myshadingD}
{5.5pt}{rgb(0pt)=(0,0,0); rgb(14pt)=(0.7,0.7,0.7 ); rgb(100pt)=(1,1,1)}
\pgftext[at=\pgfpoint{0cm}{-80pt}] {\pgfuseshading{myshadingD}}

\pgfdeclarehorizontalshading{myshadingD}
{5.5pt}{rgb(0pt)=(0,0,0); rgb(13pt)=(0.7,0.7,0.7 ); rgb(100pt)=(1,1,1)}
\pgftext[at=\pgfpoint{0cm}{-85pt}] {\pgfuseshading{myshadingD}}

\pgfdeclarehorizontalshading{myshadingD}
{5.5pt}{rgb(0pt)=(0,0,0); rgb(14pt)=(0.7,0.7,0.7 ); rgb(100pt)=(1,1,1)}
\pgftext[at=\pgfpoint{0cm}{-90pt}] {\pgfuseshading{myshadingD}}

\pgfdeclarehorizontalshading{myshadingD}
{5.5pt}{rgb(0pt)=(0,0,0); rgb(15.5pt)=(0.7,0.7,0.7 ); rgb(100pt)=(1,1,1)}
\pgftext[at=\pgfpoint{0cm}{-95pt}] {\pgfuseshading{myshadingD}}


\foreach \y in {0, -25, -50, -75} {

\foreach \x in {200, 225, 250, 275} {

\pgfdeclarehorizontalshading{myshadingD}
{4.5pt}{rgb(0pt)=(0,0,0); rgb(4pt)=(0.7,0.7,0.7 ); rgb(25pt)=(1,1,1)}
\pgftext[at=\pgfpoint{\x pt}{\y - 0pt}] {\pgfuseshading{myshadingD}}

\pgfdeclarehorizontalshading{myshadingD}
{4.5pt}{rgb(0pt)=(0,0,0); rgb(4.5pt)=(0.7,0.7,0.7 ); rgb(25pt)=(1,1,1)}
\pgftext[at=\pgfpoint{\x pt}{\y -4pt}] {\pgfuseshading{myshadingD}}

\pgfdeclarehorizontalshading{myshadingD}
{4.5pt}{rgb(0pt)=(0,0,0); rgb(6pt)=(0.7,0.7,0.7 ); rgb(25pt)=(1,1,1)}
\pgftext[at=\pgfpoint{\x pt}{\y -8pt}] {\pgfuseshading{myshadingD}}

\pgfdeclarehorizontalshading{myshadingD}
{4.5pt}{rgb(0pt)=(0,0,0); rgb(7.25pt)=(0.7,0.7,0.7 ); rgb(25pt)=(1,1,1)}
\pgftext[at=\pgfpoint{\x pt}{\y -12pt}] {\pgfuseshading{myshadingD}}

\pgfdeclarehorizontalshading{myshadingD}
{4.5pt}{rgb(0pt)=(0,0,0); rgb(7pt)=(0.7,0.7,0.7 ); rgb(25pt)=(1,1,1)}
\pgftext[at=\pgfpoint{\x pt}{\y -16pt}] {\pgfuseshading{myshadingD}}

\pgfdeclarehorizontalshading{myshadingD}
{5pt}{rgb(0pt)=(0,0,0); rgb(7.5pt)=(0.7,0.7,0.7 ); rgb(25pt)=(1,1,1)}
\pgftext[at=\pgfpoint{\x pt}{\y -20pt}] {\pgfuseshading{myshadingD}}

}

}

\color{red}

\pgfline{\pgfxy(-1.75,0.1)}{\pgfxy(1.75,0.1)}
\pgfline{\pgfxy(-1.75,-0.78)}{\pgfxy(1.75,-0.78)}
\pgfline{\pgfxy(-1.75,-1.67)}{\pgfxy(1.75,-1.67)}
\pgfline{\pgfxy(-1.75,-2.56pt)}{\pgfxy(1.75,-2.56pt)}
\pgfline{\pgfxy(-1.75,-3.44pt)}{\pgfxy(1.75,-3.44pt)}

\pgfline{\pgfxy(-1.75,0.1)}{\pgfxy(-1.75,-3.44pt)}
\pgfline{\pgfxy(-0.875,0.1)}{\pgfxy(-0.875,-3.44pt)}
\pgfline{\pgfxy(0,0.1)}{\pgfxy(0,-3.44pt)}
\pgfline{\pgfxy(0.875,0.1)}{\pgfxy(0.875,-3.44pt)}
\pgfline{\pgfxy(1.75,0.1)}{\pgfxy(1.75,-3.44pt)}

\pgfline{\pgfxy(6.6,0.1)}{\pgfxy(6.6,-3.44pt)}
\pgfline{\pgfxy(7.47,0.1)}{\pgfxy(7.47,-3.44pt)}
\pgfline{\pgfxy(8.35,0.1)}{\pgfxy(8.35,-3.44pt)}
\pgfline{\pgfxy(9.22,0.1)}{\pgfxy(9.22,-3.44pt)}
\pgfline{\pgfxy(10.1,0.1)}{\pgfxy(10.1,-3.44pt)}

\pgfline{\pgfxy(6.6,0.1)}{\pgfxy(10.1,0.1)}
\pgfline{\pgfxy(6.6,-0.76)}{\pgfxy(10.1,-0.76)}
\pgfline{\pgfxy(6.6,-1.645)}{\pgfxy(10.1,-1.645)}
\pgfline{\pgfxy(6.6,-2.53)}{\pgfxy(10.1,-2.53)}
\pgfline{\pgfxy(6.6,-3.44pt)}{\pgfxy(10.1,-3.44pt)}


\end{pgfscope}

\end{tikzpicture}

\caption{Distribution of non-zero coefficients for initial and balanced matrices}
\label{fig:balancing}
\vspace*{-0.75cm}

\end{figure}


\vspace*{-0.25cm}
 
\subsection{The partial SpMV}
\vspace*{-0.25cm}

The matrix is stored in a sparse format, adapted from the \textit{Compressed Sparse Row} (CSR) format for the particular  distribution of the non-zero coefficients.

We chose to implement the arithmetic operations in $\mathbb{Z}/\ell\mathbb{Z}$ using the \textit{Residue Number System} (RNS). The use of this representation for finite field arithmetic provides a fine grained parallelism, which can be exploited by Single Instruction, Multiple Data (SIMD) architectures.

In the remainder of the article, we consider the partial matrix--vector product as a \textit{black box}, that is, a subroutine which, on inputs $A$ and $u$ returns the product $Au$. We give more details about how this subroutine is implemented in~\cite{JELJ13}. 

\vspace*{-0.25cm}

\subsection{Communication concerns}
\vspace*{-0.25cm}

We now focus on how to share data between the computing nodes, in the cases of CPU nodes and GPU nodes.

\vspace*{-0.25cm}

\subsubsection{CPU communications}
\vspace*{-0.25cm}

The case of CPU-only setups is quite straightforward, as we use the MPI operation \texttt{MPI\_Reduce} to collect and combine on a diagonal node the results of nodes belonging to the same row, and \texttt{MPI\_Bcast} to broadcast the combined results to the nodes of each column. In the following subsections, we assume that we execute the application over a cluster of GPUs and we discuss the data movement. We restrict to NVIDIA graphics hardware. Distributing an SpMV on several GPUs requires considering two possible (and not mutually exclusive) cases: the first one where a single CPU node harbors two or more GPUs, and the second one where the GPUs are in different CPU nodes.  

\vspace*{-0.25cm}

\subsubsection{Intra-node GPU communications}

We are in the case of sharing data between two GPUs within the same CPU node. In order to do so, \textit{CUDA}, the parallel programming model for NVIDIA GPUs~\cite{CUDA} offers three possibilities:
\vspace*{-0.2cm}
\begin{itemize}
\item Staging through CPU: the communication has to involve the host CPU. Thus, it is composed of two transfers, a device-to-host copy (D2H) then a host-to-device copy (H2D).  
\item Device-to-device copy (D2D): from the programmer's perspective, it is a direct copy of the GPU buffers. Although the transfer still passes through the host memory, the copy is fully pipelined. 
\item Peer-to-Peer Direct Access (P2P DMA): using this feature, the devices can share data independently of the CPU. P2P DMA requires to enable peer access for each GPU, which is supported by recent hardware. 
\end{itemize}
\vspace*{-0.2cm}

\begin{figure}[h]

\centering
\begin{tikzpicture}[scale=0.7]

    \begin{axis}[
    		xmode=log,
        xlabel=Buffer size (kB),
        ylabel=Bandwidth (GB/s),
        ymax=15,
        ymin=-0.5,
        ymajorgrids,
        ytick = {0,2,4,6,8,10,12,14}
    ]

    \addplot [color=cyan,mark=o] plot coordinates {
        (0.1,    0.005045)
        (1,    0.050140)
        (10,    0.466853)
        (100,   2.794233)
        (1000,   5.549174)
        (10000,   6.155752)
        (100000,  6.085420)
        (1000000,  6.07)
    };

	\addplot [color=magenta,mark=square*]plot coordinates {
        (0.1,     0.005117)
        (1,       0.049388)
        (10,      0.412660)
        (100,     2.634907)
        (1000,    4.459269)
        (10000,   6.758679)
        (100000,  7.318176)
        (1000000,  7.32)
    };	
	
	\addplot plot coordinates {};

	\addplot plot coordinates {
        (0.1,     0.006939)
        (1,       0.069161)
        (10,      0.669837)
        (100,     4.430856)
        (1000,    9.147206)
        (10000,   10.248327)
        (100000,  10.386247)
        (1000000,  10.377501)
    };		

    \legend{D2H + H2D\\D2D\\D2D (P2P DMA)\\}

    \end{axis}
    
\draw [thick] (8,4.5) rectangle (9.25,5.25);
\node[right] at (7.875,4.875){\textbf{\scriptsize GPU$_0$}};

\draw [thick] (11,4.5) rectangle (12.25,5.25);
\node[right] at (10.875,4.875){\textbf{\scriptsize GPU$_1$}};

\draw [thick] (8,2) rectangle (12.25,2.5);
\node[right] at (9,2.25){\textbf{Chipset}};
\draw [thick] (8,1.5) rectangle (12.25,0.25);
\node[right] at (8.5,1){\textbf{Host Memory}};

\draw [fill=gray] (8.5,4.5) rectangle (8.75,2.5);
\draw [fill=gray ] (11.5,4.5) rectangle (11.75,2.5);

\draw [fill=gray ] (9.8,2) rectangle (10.3,1.5);

\node[right] at (8.6,3.5){\textit{\footnotesize PCIe}};
\node[right] at (10.2,3.5){\textit{\footnotesize PCIe}};

\end{tikzpicture}

\caption{Benchmarking Intra-node GPU communications}
\label{fig:intra-node GPU bench}
\vspace*{-0.5cm}

\end{figure}
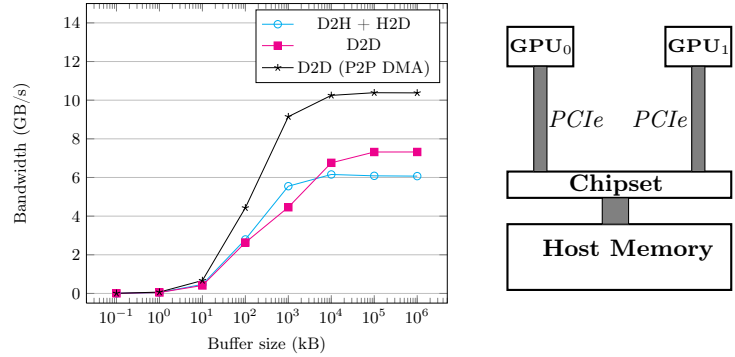

The P2P DMA feature should decrease the host overhead and thus accelerate the memory copies. To verify it, we ran benchmarks to compare the bandwidth and latency of each approach (cf. Figure~\ref{fig:intra-node GPU bench}). The experiment is performed using two NVIDIA GeForce GTX 680 cards. The benchmarks measure the run time for sending messages of increasing size from one GPU to the other. The latencies for the first two options are \SI{19.7}{\micro\second} and \SI{19.4}{\micro\second}, respectively, and only \SI{14}{\micro\second} when the P2P DMA is enabled. The peak bandwidths are 6.1~GB/s for the explicit host staging transfer, 7.3~GB/s for the device to device transfer, and 10.4~GB/s for the P2P DMA transfer.

\vspace*{-0.5cm}

\subsubsection{Inter-node GPU communications} 

Now, we are interested in the case of sharing data between GPUs installed in different CPU nodes. The trivial option in this case is to perform the transfer in three steps: a data copy from device to host using CUDA routines, then use MPI to copy data between hosts, and finally a CUDA copy from host to device on the destination node (cf. Figure~\ref{fig:without Cuda-aware}). 

It is however possible to overcome the host staging using the \textit{Cuda-aware MPI} feature which combines MPI and CUDA. It allows one to address GPU buffers directly in the MPI routines (cf. Figure~\ref{fig:with Cuda-aware}). From the programmer's point of view, a data transfer boils down to one call to an MPI routine. With \textit{Cuda-aware MPI}, the data transfers are fully pipelined, while without the feature, the transfers between hosts and those between the device and the host are pipelined separately. The \textit{Cuda-aware MPI} feature is incorporated in several widely used MPI libraries and considerably improves the data movement latencies.

\vspace*{-0.5cm}

\begin{figure}[h]

\centering
\begin{tikzpicture}[scale=0.7]

\draw [thick] (0,0) rectangle (1.25,0.75);
\node[right] at (-0.1,0.325){\textbf{\scriptsize GPU$_0$}};

\draw [thick] (11,0) rectangle (12.25,0.75);
\node[right] at (10.9,0.325){\textbf{\scriptsize GPU$_1$}};

\draw [thick] (2.75,0) rectangle (3.5,0.5);
\node[right] at (2.75,0.25){\textbf{\scriptsize IB}};

\draw [thick] (8.75,0) rectangle (9.5,0.5);
\node[right] at (8.75,0.25){\textbf{\scriptsize IB}};

\draw [thick] (0,-2) rectangle (3.5,-1.5);
\node[right] at (1,-1.75){\textbf{\scriptsize Chipset}};

\draw [thick] (8.75,-2) rectangle (12.25,-1.5);
\node[right] at (9.55,-1.75){\textbf{\scriptsize Chipset}};

\draw [thick] (0,-2.5) rectangle (3.5,-4.25);
\node[right] at (0.25,-4){\textbf{\scriptsize Host$_0$ Memory}};

\draw [thick] (8.75,-2.5) rectangle (12.25,-4.25);
\node[right] at (9,-4){\textbf{\scriptsize Host$_1$ Memory}};

\draw [thick] (0.25,-2.75) rectangle (1.5,-3.625);
\node[right] at (0.2,-3){\scriptsize CUDA};
\node[right] at (0.2,-3.375){\scriptsize buffer};

\draw [thick] (10.75,-2.75) rectangle (12,-3.625);
\node[right] at (10.7,-3){\scriptsize CUDA};
\node[right] at (10.7,-3.375){\scriptsize buffer};

\draw [thick] (2.25,-2.75) rectangle (3.35,-3.625);
\node[right] at (2.2,-3){\scriptsize CPU};
\node[right] at (2.2,-3.375){\scriptsize buffer};

\draw [thick] (8.9,-2.75) rectangle (10,-3.625);
\node[right] at (8.8,-3){\scriptsize CPU};
\node[right] at (8.8,-3.375){\scriptsize buffer};

\draw [fill=gray ] (0.5,0) rectangle (0.75,-1.5);
\draw [fill=gray ] (3,0) rectangle (3.25,-1.5);

\draw [fill=gray ] (11.5,0) rectangle (11.75,-1.5);
\draw [fill=gray ] (9,0) rectangle (9.25,-1.5);

\draw [fill=gray ] (1.6,-2) rectangle (2.1,-2.5);
\draw [fill=gray ] (10.3,-2) rectangle (10.8,-2.5);

\draw [fill=gray ] (3.5,0.15) rectangle (8.75,0.3);

\node[right] at (-0.6,-0.75){\textit{\scriptsize PCIe}};
\node[right] at (3.2,-0.75){\textit{\scriptsize PCIe}};
\node[right] at (5.7,-0.25){\textit{\scriptsize IB}};

\draw [->][ultra thick][blue](0.9,0)--(0.9,-2.75);
\draw [->][ultra thick][blue](9.35,0)--(9.35,-2.75);

\draw [->][ultra thick](1.5,-3.25)--(2.25,-3.25);
\draw [->][ultra thick](10,-3.25)--(10.75,-3.25);

\draw [->][ultra thick][blue](2.9,-2.75)--(2.9,0);
\draw [->][ultra thick][blue](11.35,-2.75)--(11.35,0);

\draw [->][ultra thick][red](3.5,0.4)--(8.75,0.4);

\draw [dashed][orange] (2.75,-2.25) -- (2.75,-5.25);
\draw [dashed][orange] (9.5,-2) -- (9.5,-5);
\node[right] at (0.5,-4.75){\textit{\scriptsize GPU to}};
\node[right] at (0.5,-5.12){\textit{\scriptsize CPU copy}};
\node[right] at (4.25,-4.75){\textit{\scriptsize CPU to CPU transfer}};
\node[right] at (9.75,-4.75){\textit{\scriptsize CPU to}};
\node[right] at (9.75,-5.12){\textit{\scriptsize GPU copy}};

\end{tikzpicture}

\caption{Data copy from GPU$_0$ to GPU$_1$ without \textit{Cuda-aware MPI}} 
\label{fig:without Cuda-aware}

\vspace*{-0.75cm}

\end{figure}
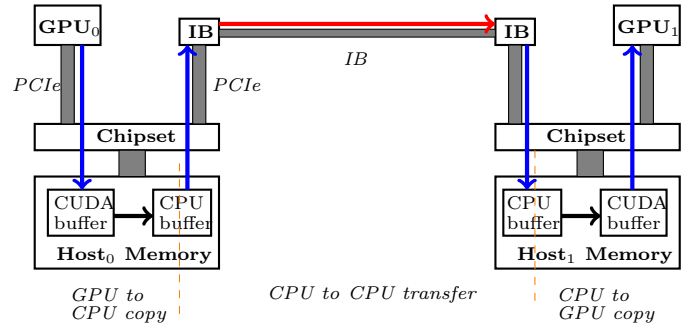


\begin{figure}[h]

\centering

\begin{tikzpicture}[scale=0.7]

\draw [thick] (0,0) rectangle (1.25,0.75);
\node[right] at (-0.1,0.325){\textbf{\scriptsize GPU$_0$}};

\draw [thick] (11,0) rectangle (12.25,0.75);
\node[right] at (10.9,0.325){\textbf{\scriptsize GPU$_1$}};

\draw [thick] (2.75,0) rectangle (3.5,0.5);
\node[right] at (2.75,0.25){\textbf{\scriptsize IB}};

\draw [thick] (8.75,0) rectangle (9.5,0.5);
\node[right] at (8.75,0.25){\textbf{\scriptsize IB}};

\draw [thick] (0,-2) rectangle (3.5,-1.5);
\node[right] at (1,-1.75){\textbf{\scriptsize Chipset}};

\draw [thick] (8.75,-2) rectangle (12.25,-1.5);
\node[right] at (9.55,-1.75){\textbf{\scriptsize Chipset}};

\draw [thick] (0,-2.5) rectangle (3.5,-4.25);
\node[right] at (0.25,-4){\textbf{\scriptsize Host$_0$ Memory}};

\draw [thick] (8.75,-2.5) rectangle (12.25,-4.25);
\node[right] at (9,-4){\textbf{\scriptsize Host$_1$ Memory}};

\draw [thick] (0.25,-2.75) rectangle (1.5,-3.625);
\node[right] at (0.2,-3){\scriptsize CUDA};
\node[right] at (0.2,-3.375){\scriptsize buffer};

\draw [thick] (10.75,-2.75) rectangle (12,-3.625);
\node[right] at (10.7,-3){\scriptsize CUDA};
\node[right] at (10.7,-3.375){\scriptsize buffer};

\draw [thick] (2.25,-2.75) rectangle (3.35,-3.625);
\node[right] at (2.2,-3){\scriptsize CPU};
\node[right] at (2.2,-3.375){\scriptsize buffer};

\draw [thick] (8.9,-2.75) rectangle (10,-3.625);
\node[right] at (8.8,-3){\scriptsize CPU};
\node[right] at (8.8,-3.375){\scriptsize buffer};

\draw [fill=gray ] (0.5,0) rectangle (0.75,-1.5);
\draw [fill=gray ] (3,0) rectangle (3.25,-1.5);

\draw [fill=gray ] (11.5,0) rectangle (11.75,-1.5);
\draw [fill=gray ] (9,0) rectangle (9.25,-1.5);

\draw [fill=gray ] (1.6,-2) rectangle (2.1,-2.5);
\draw [fill=gray ] (10.2,-2) rectangle (10.7,-2.5);

\draw [fill=gray ] (3.5,0.15) rectangle (8.75,0.3);

\node[right] at (-0.6,-0.75){\textit{\scriptsize PCIe}};
\node[right] at (3.2,-0.75){\textit{\scriptsize PCIe}};
\node[right] at (5.7,-0.25){\textit{\scriptsize IB}};

\draw [->][ultra thick][blue](0.9,0)--(0.9,-2.75);
\draw [->][ultra thick][blue](9.35,0)--(9.35,-1.9)--(10.875,-1.9)--(10.875,-2.75);

\draw [->][ultra thick][blue](1.4,-2.75)--(1.4,-1.9)--(2.9,-1.9)--(2.9,0);
\draw [->][ultra thick][blue](11.35,-2.75)--(11.35,0);

\draw [->][ultra thick][red](3.5,0.4)--(8.75,0.4);

\end{tikzpicture}

\caption{Data copy from GPU$_0$ to GPU$_1$ with \textit{Cuda-aware MPI}} 
\label{fig:with Cuda-aware}

\vspace*{-0.75cm}

\end{figure}


In Figure~\ref{fig:inter-node GPU bench}, we report the results of bandwidth benchmarks for inter-node GPU-to-GPU communications. We ran the experiment using two NVIDIA GTX 680 installed in two nodes connected with QDR InfiniBand. We use CUDA 5.0 and Open MPI 1.7.3. In addition to  benchmarks for the two ways of communication, we added the Host-to-Host (H2H) communication results as a reference, for which we measured the data movement from one CPU buffer to another CPU buffer using the regular MPI routines.

The latency of a plain Device-to-Device transfer is \SI{11}{\micro\second}. It becomes \SI{9}{\micro\second} if the feature \textit{Cuda-aware MPI} is used. The latency of the Host-to-Host transfer is \SI{1}{\micro\second}. Without \textit{Cuda-aware MPI}, the bandwidth is bounded by 2.3 GB/s. The \textit{Cuda-aware MPI} feature allows to reach the Host-to-Host peak bandwidth, which is 3.7 GB/s. 

\vspace*{-0.75cm}

\begin{figure}[h]

\centering
\begin{tikzpicture}[scale=0.75]

    \begin{axis}[
    		xmode=log,
        xlabel=Buffer size (kB),
        ylabel=Bandwidth (GB/s),
        ymax=6,
        ymin=-0.5,
        ymajorgrids,
        ytick = {0,1,2,3,4,5,6,7}
    ]

	\addplot [color=cyan,mark=o] plot coordinates {
		(0.1,    0.007580)
        (1,    	0.079783)
        (10,    0.653595)
        (100,   1.729087)
        (1000,   2.266114)
        (10000,   2.316787)
        (100000,  2.282138)
        (1000000, 2.311904)
    };	

	\addplot [color=magenta,mark=square*]plot coordinates {
		(0.1,    0.003827)
        (1,    	0.083015)
        (10,    0.777968)
        (100,   1.289790)
        (1000,   3.133873)
        (10000,   3.661893)
        (100000,  3.731204)
        (1000000,  3.740043)
    };			

	\addplot plot coordinates {};

    \addplot[color=black,mark=x] plot coordinates {
        (0.1,    0.060827)
        (1,    	0.600240)
        (10,    0.063513)
        (100,   3.065228)
        (1000,   3.742431)
        (10000,   3.696899)
        (100000,  3.654803)
        (1000000,  3.646256)
    };

    \legend{D2D (without Cuda-aware MPI)\\D2D (with Cuda-aware MPI)\\H2H\\}

    \end{axis}

\end{tikzpicture}

\caption{Benchmarking Inter-node GPU communications}
\label{fig:inter-node GPU bench}
\vspace*{-0.75cm}

\end{figure}	


Another feature that further optimizes data transfers is \textit{GPU Direct}. The \textit{GPU Direct} offers lower latency for moving data compared to transfers staged through the host. However, its bandwidth is significantly limited. We could not deploy this feature in our application, as it is supported only by the recent Tesla and Quadro cards. A comparison of the performance of this feature with the transfers staged through host can be found in~\cite{POTL13}.

\vspace*{-0.25cm}
\section{Examples of computations}
\label{sec:examples of computation}
\vspace*{-0.25cm}
\subsection{DLP in GF$(2^{809})^\times$ using FFS}
\vspace*{-0.25cm}
The function field sieve (FFS)~\cite{ADLE94} is an \textit{index-calculus} algorithm designed to attack the DLP in the multiplicative subgroup of a finite field GF($p^n$), where the characteristic $p$ is a small prime. Barbulescu et al.~announced in \cite{FFS809} the solving of the DLP in the 202-bit prime order subgroup of GF$(2^{809})^\times$ using FFS. This computation is the largest DLP computation in a binary field extension of prime degree. The previous record was the computation of a DLP in GF$(2^{613})^\times$~\cite{JOUX05}.

\vspace*{-0.45cm}

\subsubsection{Matrix.}
In this computation, the linear algebra step is performed in $\mathbb{Z}/\ell\mathbb{Z}$ where $\ell$ is 202 bits long. The relation collection phase produced an initial matrix of 78.8M rows. A preliminary structured Gaussian elimination reduced the matrix to 3,602,667 rows and columns, with an average of 100 non-zero coefficients per row. Each non-zero coefficient of $A$ fits in a single machine word. Around 90\% of them are $\pm1$ \cite{Bouvier12}, \cite{FFS809}.

\vspace*{-0.45cm}

\subsubsection{Linear Algebra Setup.}
	At the time of the computation, we had access to a 4-node cluster, with 2.4~GHz Intel Xeon E5620 Westmere processors connected with InfiniBand network at~40 Gb/s. Each node is equipped with 2 NVIDIA Tesla M2050 graphics processors.
	
	The total memory required to handle the matrix along with the input and output vectors is 3.16~GB. Since the available memory on one card is only 3~GB, the block Widemann configuration $(n=8,m=16)$, for which a sequence $\left (^txA^iy^{(j)}\right )_{i \in \mathbb{N}}$ can be computed on a single device, is not feasible. We have to compute each sequence on more than one device; the configuration $(n=4,m=8)$ with a $2 \times 1$ split of the matrix and the configuration $(n=2,m=4)$ with a $2 \times 2$ split of the matrix are possible. Theoretically, the former appears to be the most convenient, since only two GPUs connected to the same node communicate, while, with the latter, 4 GPUs interconnected with the network are required to communicate.
	
In the following table, we detail a comparison between these two configurations. The comparison shows how the inter-node GPU communication for the second configuration slows down the overall computation time. We also present benchmarks related to a smaller matrix, for which the three configurations are possible and a bigger matrix, for which only the $(n=2,m=4)$ configuration is feasible. 
	
	\vspace*{-0.2cm}

\begin{center}
    \begin{tabular}{|c||c|c|c|c|}
      \hline
      \small Matrix size & \small Possible blocking & \small SpMV + comm. & \small Overall & \small Ratio com.\\
      \small (required memory) & \small parameters & \small delays per iteration & \small comp. time & \small /iteration\\
      \hline
	  \footnotesize 3.6M $\times$ 3.6M & \footnotesize \boldmath $(n=4,m=8)$ & \bf \footnotesize 142 + 27 ms & \bf \footnotesize 4.5 days & \bf \footnotesize 16\%\\
	  \footnotesize (3.2 GB) & \footnotesize $(n=2,m=4)$ & \footnotesize 72 + 41 ms & \footnotesize 6 days & \footnotesize 37\%\\
	  \hline
      \multirow{2}{*}{\footnotesize 3M $\times$ 3M} & \footnotesize \boldmath  $(n=8,m=16)$ & \bf \footnotesize 228 + 0 ms & \bf \footnotesize 2.5 days & \bf \footnotesize 0\%\\
      \multirow{2}{*}{\footnotesize (2.7 GB)}& \footnotesize $(n=4,m=8)$ & \footnotesize 115 + 23 ms & \footnotesize 3 days & \footnotesize 17\%\\
      & \footnotesize $(n=2,m=4)$ & \footnotesize 58 + 35 ms & \footnotesize 4.1 days & \footnotesize 38\%\\
	  \hline
	  \footnotesize 6M $\times$ 6M & \multirow{2}{*}{\footnotesize \boldmath $(n=2,m=4)$} & \multirow{2}{*}{\bf \footnotesize 123 + 69 ms} &  \multirow{2}{*}{\bf \footnotesize 16.7 days} & \multirow{2}{*}{\bf \footnotesize 36\%}\\
	  \footnotesize (5.4 GB) &  &  &  & \\
      \hline
    \end{tabular}
  \label{FFS-809}
\end{center}
	\vspace*{-0.2cm}

	With the $(n=4,m=8)$ blocking parameters, an iteration takes 169~ms on each node, including 27~ms for the GPU communications. The initial sequence computation required 2.6 days in parallel on the 4 nodes. The linear generator computation was carried out in parallel using 16 threads running on 16 CPU cores. It required 2 hours. Finally, computing the kernel vector required 1.8 days in parallel on the 4 GPU nodes. The overall computation took a total wall-clock time of about 4.5 days.     
 
\vspace*{-0.25cm}

\subsection{DLP in a 596-bit prime field using NFS}
To compute discrete logarithms in a prime field GF($p$), the Number Field Sieve (NFS) algorithm is used~\cite{JoLe03}. The last NFS record was accomplished by T. Kleinjung et al.~\cite{BAHR07} for a 530-bit (160-decimal-digit) prime $p$ using NFS. We are currently running an NFS-based computation to attack the DLP in a 596-bit (180-decimal-digit) prime field. The linear algebra step is defined over a 595-bit prime $\ell$.

\vspace*{-0.45cm}

\subsubsection{Matrix.}
The matrix contains 179M rows at the end of the relation collection. The preliminary structured Gaussian elimination reduced the number of rows to 7,287,476, with an average weight of 150 non-zero coefficients per row. The matrices issued from NFS computations contain a small number (5, here) of dense columns, whose elements live in $\mathbb{Z}/\ell\mathbb{Z}$. The rest of the matrix is similar to an FFS matrix in terms of distribution and coefficient size. Taking this dense part into account adds a non-negligible cost when compared to FFS matrices.

\vspace*{-0.45cm}

\subsubsection{GPU Setup.}
For this computation, we have access to 8 NVIDIA GeForce GTX 680 graphics processors, plugged into a 4-node cluster of Intel Xeon E5-2609 processors running at~2.4 GHz, and connected with QDR Infiniband network. Each graphics card has 4 GB of memory. The total memory required to carry out the SpMV on one GPU is 9.8 GB. Thus, 4 GPUs should work on a single sequence, \textit{i.e.}, at most two sequences can be computed in parallel, and the blocking parameters are $(n=2,m=4)$. An iteration takes 615~ms on each group of 4 GPUs, with 195~ms for the GPU-to-GPU communications. The overall computation should take a total wall-clock time of about 65 days.

\vspace*{-0.45cm}

\subsubsection{CPU Setup.}
Another option was tried, using our CPU implementation on a 768-core cluster. The cluster contains 48 nodes connected with FDR Infiniband. Each node hosts two 2-GHZ 8-core Intel Xeon E5-2650 processors. With this setup, we propose an 8$\times$8 split of the matrix, so that 64 MPI processes running on 4 nodes work together to carry out a matrix--vector product, each process running on one core. This yields to a $(n=12,m=24)$ block Wiedemann configuration. The processes are distributed so that the processes running on the same node are contiguous. This allows to accelerate the reduction/broadcast operations, since data sharing between threads belonging to the same node is performed on shared memory, not across the network. A speedup of 2.4 on the communication delay is observed when comparing with the default MPI processes mapping.

In the following table, we compare the GPU and CPU setups. We observe that starting from a certain matrix size and with these setups, the multi-core acceleration prevails over the GPU one. For comparison, we add a setup, where we have a cluster similar to the 48-node multi-core cluster, but containing two NVIDIA GeForce GTX 680 on each node. This setup is not speculative, the GPU setup obtained with 8 GPUs scales perfectly to 96 GPUs, thanks to the cost-free distribution of block Wiedemann algorithm.  

	\vspace*{-0.5cm}

\begin{center}
    \begin{tabular}{|c|c||c|c|c|c|}
      \hline
      \small Matrix size & \multirow{2}{*}{\small Setup} & \small Blocking & \small SpMV+com. & \small Overall & \small Ratio \\
       \small (Memory) & & \small parameters & \small delays [ms] & \small comp. time & \small com.\\
      \hline
	  \multirow{5}{*}{\footnotesize	7.3M $\times$ 7.3M} & \footnotesize 8 GPUs & \footnotesize $(n=2,m=4)$ & \multirow{2}{*}{\footnotesize 420 + 195} & \multirow{2}{*}{\footnotesize 65 days} & \multirow{2}{*}{\footnotesize 32\%}\\      
	  \multirow{5}{*}{\footnotesize (9.8 GB)} & \footnotesize on 4 nodes & \footnotesize 4 GPUs $\leftrightarrow$ 1 subtask &  &  & \\
	  \cline{2-6}      
      & \footnotesize \bf 768 cores & \footnotesize \boldmath $(n=12,m=24)$ & \multirow{2}{*}{\footnotesize \bf 1700 + 400} & \multirow{2}{*}{\footnotesize \bf 39 days} & \multirow{2}{*}{\footnotesize \bf 19\%}\\      
	   & \footnotesize \bf on 48 nodes & \bf \footnotesize 64 cores $\leftrightarrow$ 1 subtask  &  &  & \\
   	  \cline{2-6}      
      & \footnotesize 96 GPUs & \footnotesize $(n=24,m=48)$ & \multirow{2}{*}{\footnotesize 420 + 195} & \multirow{2}{*}{\footnotesize 5.5 days} & \multirow{2}{*}{\footnotesize 32\%}\\      
	   & \footnotesize on 48 nodes & \footnotesize 4 GPUs $\leftrightarrow$ 1 subtask  &  &  & \\      
      \hline
    \end{tabular}
  \label{NFS-596}
\end{center}

	\vspace*{-0.25cm}

With the CPU setup, an iteration is performed in 2.1~s by the 64 parallel threads, including 0.4~ms for communications. The \textit{Krylov} phase required 22 in the 768-core cluster, which is equivalent to 46-core years. The \textit{Lingen} phase required 15 hours running on 144 cores. The \textit{Mksol} phase took 16 days in the 768-core cluster (i.e. 34-core years). The overall computation required around 80-core years.

\vspace*{-0.5cm}

\section{Conclusion}
\label{sec:conclusion}
	\vspace*{-0.2cm}

In this article, we presented how the block solvers, in our case block Wiedemann algorithm, distribute heavy computations without an additional overhead. We discussed a further parallelization of the matrix-vector product and detailed how we can efficiently run this computation in a cluster of GPUs or CPUs. In the examples that we ran, we did not combine the two architectures on the same computation. However, our final implementation can be run on a hybrid GPU/Multi-core architecture.  

\vspace*{-0.25cm}

\subsubsection{Acknowledgments}

The author is grateful to Jérémie Detrey and Emmanuel Thomé. This work would not be possible without their support. We also thank the reviewers for their valuable comments.


\vspace*{-0.25cm}

\bibliographystyle{plain}
\bibliography{linalg}

\end{document}